\documentclass[sigconf]{acmart}

\AtBeginDocument{%
  }

\usepackage{subcaption}
\usepackage{algorithmic}
\usepackage{graphicx}
\usepackage{fancyvrb}
\usepackage{listings}
\usepackage{makecell}
\usepackage{multirow}

\copyrightyear{2026}
\acmYear{2026}
\setcopyright{cc}
\setcctype{by}
\acmConference[AST '26]{7th ACM/IEEE International Conference on Automation of Software Test (AST 2026)}{April 13--14, 2026}{Rio de Janeiro, Brazil}
\acmBooktitle{7th ACM/IEEE International Conference on Automation of Software Test (AST 2026) (AST '26), April 13--14, 2026, Rio de Janeiro, Brazil}
\acmPrice{}
\acmDOI{10.1145/3793654.3793749}
\acmISBN{979-8-4007-2476-3/2026/04}

\begin{document}

\title{Testing Framework Migration with Large Language Models}

\author{Altino Alves}
\orcid{0009-0007-2709-5950}
\affiliation{%
  \institution{DCC, UFMG}
  \city{Belo Horizonte}
  \country{Brazil}}
\email{altinojunior@dcc.ufmg.br}

\author{João Eduardo Montandon}
\orcid{0000-0002-3371-7353}
\affiliation{%
  \institution{DCC, UFMG}
   \city{Belo Horizonte}
  \country{Brazil}}
\email{joao@dcc.ufmg.br}

\author{Andre Hora}
\orcid{0000-0003-4900-1330}
\affiliation{%
  \institution{DCC, UFMG}
   \city{Belo Horizonte}
  \country{Brazil}}
\email{andrehora@dcc.ufmg.br}

\begin{abstract}
Python developers rely on two major testing frameworks: \texttt{unittest} and \texttt{Pytest}. 
While \texttt{Pytest} offers simpler assertions, reusable fixtures, and better interoperability, migrating existing suites from \texttt{unittest} remains a manual and time-consuming process.
Automating this migration could substantially reduce effort and accelerate test modernization.
In this paper, we investigate the capability of Large Language Models (LLMs) to automate test framework migrations from \texttt{unittest} to \texttt{Pytest}.
We evaluate GPT 4o and Claude Sonnet 4 under three prompting strategies (Zero-shot, One-shot, and Chain-of-Thought) and two temperature settings (0.0 and 1.0).
To support this analysis, we first introduce a curated dataset of real-world migrations extracted from the top 100 Python open-source projects.
Next, we actually execute the LLM-generated test migrations in their respective test suites.
Overall, we find that 51.5\% of the LLM-generated test migrations failed, while 48.5\% passed.
The results suggest that LLMs can accelerate test migration, but there are often caveats.
For example, Claude Sonnet 4 exhibited more conservative migrations (e.g., preserving class-based tests and legacy \texttt{unittest} references), while GPT-4o favored more transformations (e.g., to function-based tests).
We conclude by discussing multiple implications for practitioners and researchers.
\end{abstract}





\maketitle

\section{Introduction}

Software testing is fundamental to avoiding regressions and catching bugs.
Currently, Python developers can rely on two main testing frameworks: unittest~\cite{unittest} and Pytest~\cite{pytest}.
Pytest provides some advantages compared to unittest, including simpler assertions, reuse of fixtures, and interoperability~\cite{pytest, barbosa2022}.

\begin{figure}[t]
     \centering
         \includegraphics[width=0.47\textwidth]{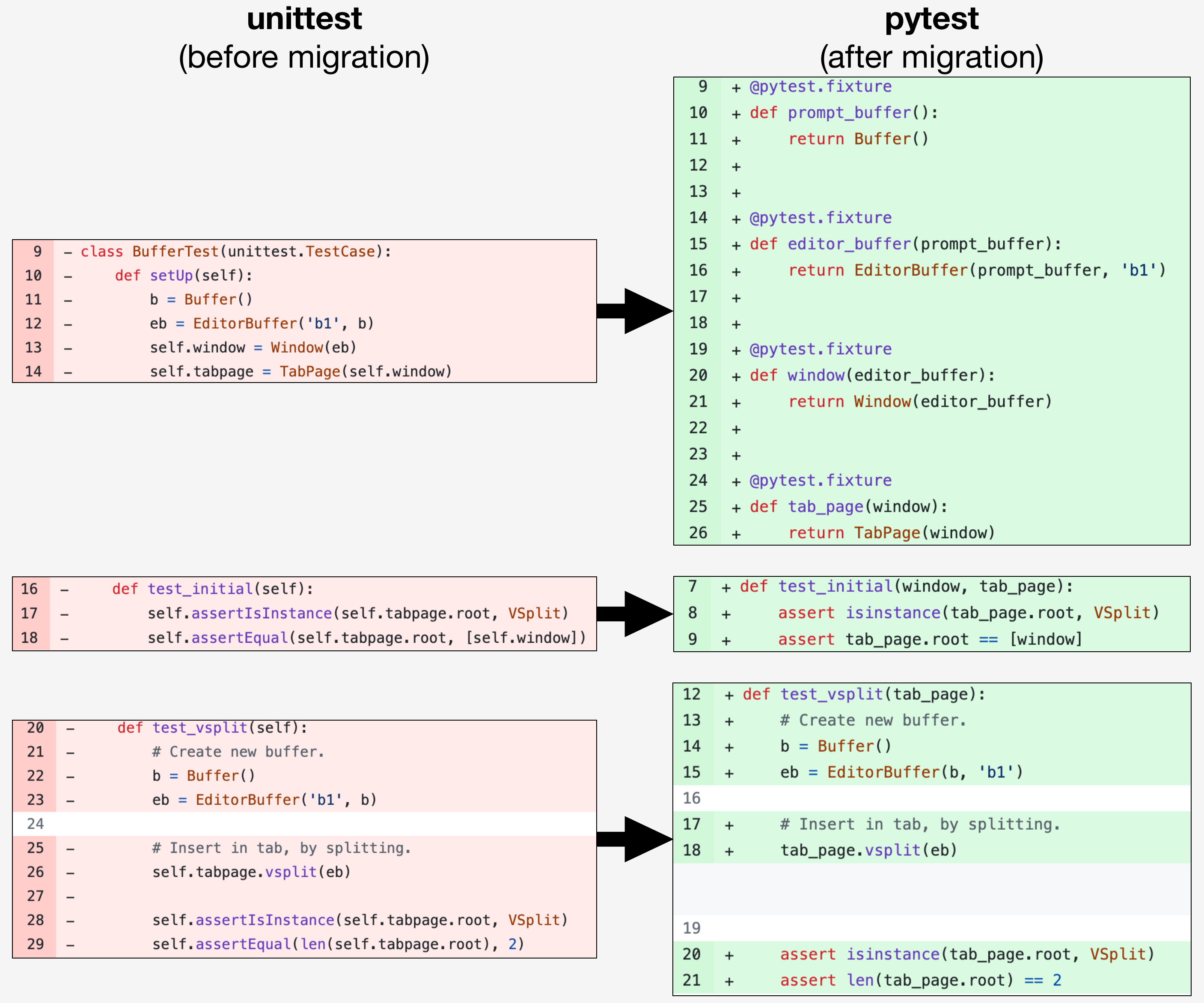}
         \caption{Migration from unittest to Pytest (pyvim).}
         \Description{Migration from unittest to Pytest (pyvim).}
        \label{fig:migration1}
\end{figure}

Multiple Python projects---such as Pandas, NumPy, ScikitLearn, Requests, and Flask---have migrated to or are currently migrating to Pytest due to these benefits~\cite{barbosa2022}.
For example, Figure~\ref{fig:migration1} presents a migration from unittest to Pytest in pyvim.\footnote{\url{https://github.com/prompt-toolkit/pyvim/commit/7e1c7bfb505cefba468}}
This migration consists of three major modifications.
The first chunk shows that the unittest \texttt{setUp} method is being split into four \texttt{@Pytest.fixtures} properties.
The second and third chunks modify test methods to receive these fixtures and use \texttt{assert} statements during the test verification.
Section~\ref{sec:background} explains this migration in more detail.

Pytest allows projects to run tests written in unittest, thus enabling a gradual migration process. 
\textbf{However, this migration can be time-consuming given the complexity of test suites: a recent study found that some projects may take months or years to conclude the migration or simply never conclude it~\cite{barbosa2022}}.
Besides, using two different testing frameworks at the same time can increase the effort needed to maintain the test suite.
Such projects could benefit from an automated solution to assist developers in migrating these tests faster and effectively.

Recently, Large Language Models (LLMs) have been evaluated in multiple software engineering tasks, including source code and tests generation, bug fix, code smells detection, and to support the code review process~\cite{openai2023gpt4, fan2023large, monteiro2023end, liang2023can, tufano2023predicting, georgsen2023beyond, hou2023large, hora2024predicting, silva2024detecting, di2025deepmig}.
Other studies investigated the use of LLMs to support API migration~\cite{esem2024_api_migration_llm, islam2025empiricalstudypythonlibrary}.
However, to the best of our knowledge, no study has explored the migration of the test frameworks.

In this paper, we investigate the capability of LLMs to assist developers during the migration of test frameworks.
For this purpose, we built a dataset with over 900 real-world migrations from \texttt{unittest} to \texttt{Pytest}, manually implemented by developers~\cite{alves2025testmigrationsinpy}.
Next, we selected a subset of migrated tests that still exist by September~2025, and thus remain executable.
This process resulted in a final set of 40 isolated migrations that are investigated in this paper.
In our experiment, we rely on GPT-4o and Claude Sonnet 4 to migrate Python tests from \texttt{unittest} to \texttt{Pytest}.
We applied three prompting strategies—Zero-shot, One-shot, and Chain-of-Thought, under two temperature settings (0.0 and 1.0), generating a total of 480 migration variants.

To ensure the migration proposed by the LLMs was correct, we reproduced the test environment of each migration and actually \emph{executed} both LLM-migrated and developer-migrated tests.
We then compared the results of both executions in terms of test correctness (i.e.,~test was successfully executed) and test coverage (i.e.,~test coverage remained the same).
Both models achieved an overall correctness rate of 48.54\%, while successful migrations maintained the same test coverage, indicating that LLMs did not modify runtime behaviour.
However, 25 out of 40 migrations presented at least one failing configuration (247 total failures), mostly related to dependency handling, fixture adaptation, or setup inconsistencies. 
While Claude Sonnet 4 preserved class-based structures and \texttt{unittest} patterns, GPT-4o favored function-based rewrites aligned with \texttt{Pytest}, revealing distinct yet complementary migration styles.
We conclude by discussing implications for researchers and practitioners.

\noindentparagraph{Contribution.}
The contributions of this paper are twofold:
(1) we provide the first study to explore test migration with LLMs and 
(2) we discuss multiple implications for researchers and practitioners.

\section{Background and Motivation}
\label{sec:background}

Unittest~\cite{unittest} and Pytest~\cite{pytest} are the most popular testing frameworks in Python~\cite{jetbrains-survey}.
Unittest belongs to the Python standard library, while Pytest is a third-party testing framework.
Unittest relies on classes and inheritance to create tests (i.e.,~the test class needs to extend the unittest class \texttt{TestCase}), whereas Pytest tests can be regular functions, with the \emph{test} prefix.
Consequently, Pytest tests tend to be less verbose than unittest ones.
Another difference is the assertions: unittest provides \texttt{self.assert*} methods (e.g.,~\texttt{assertEqual}, \texttt{assertTrue}, etc.), while Pytest allows developers to use the regular Python \texttt{assert} statement for verifying expectations and values.
There are many other differences; for example, Pytest facilitates the creation of parameterized tests and the reuse of fixtures.

Due to the advantages of Pytest, many Python projects have migrated to this framework.
A prior study discovered that 27\% of top-100 most popular Python projects migrated or were migrating to Pytest~\cite{barbosa2022}.
To gain more insights into the relevance of this problem, we replicated this study in the same set of popular projects, found that the migration rate has increased to 37\%.

Migrating from unittest to Pytest may involve at least the following major steps: 
(1) removing test from class and moving to regular functions; 
(2) replacing assertions with Pytest asserts; and
(3) moving setup/teardown operations to Pytest fixtures.
Steps 1 and 2 is relatively simple to apply because the migration between unittest and Pytest is almost direct.
For example, the test \texttt{test\_vsplit} in Figure~\ref{fig:migration1} only replaces the unittest assertions \texttt{assertIsInstance} and \texttt{assertEqual} by the Pytest \texttt{assert} statement.

On the other hand, migrating the remaining steps is harder to accomplish because there is no direct mapping between unittest and Pytest.
For example, the unittest \texttt{setUp} method is split into four Pytest fixture functions in Figure~\ref{fig:migration1}: \texttt{prompt\_buffer}, \texttt{editor\_buffer}, \texttt{window}, and \texttt{tab\_page}.
The two tests (\texttt{test\_\-initial} and \texttt{test\_\-vsplit}) are then adapted to receive the fixtures via parameters in Pytest.
When Pytest runs a test, it looks at the parameters of the test function and then searches for fixtures with the same names as those parameters~\cite{pytest, barbosa2022}.
Once Pytest finds them, it runs those fixtures, captures what they returned, and passes those objects into the test function as arguments.

\begin{center}
\fcolorbox{black}{gray!15}{
  \parbox{\dimexpr1\linewidth-2\fboxsep-2\fboxrule}{
The number of popular Python projects that decided to migrate from unittest to Pytest increased from 27\% to 37\%, reinforcing the need for an automated solution to assist this migration process.
  }
}
\end{center}


\section{Study Design}

\subsection{Overview}

\begin{figure*}[ht]
  \centering
  \includegraphics[width=0.9\textwidth]{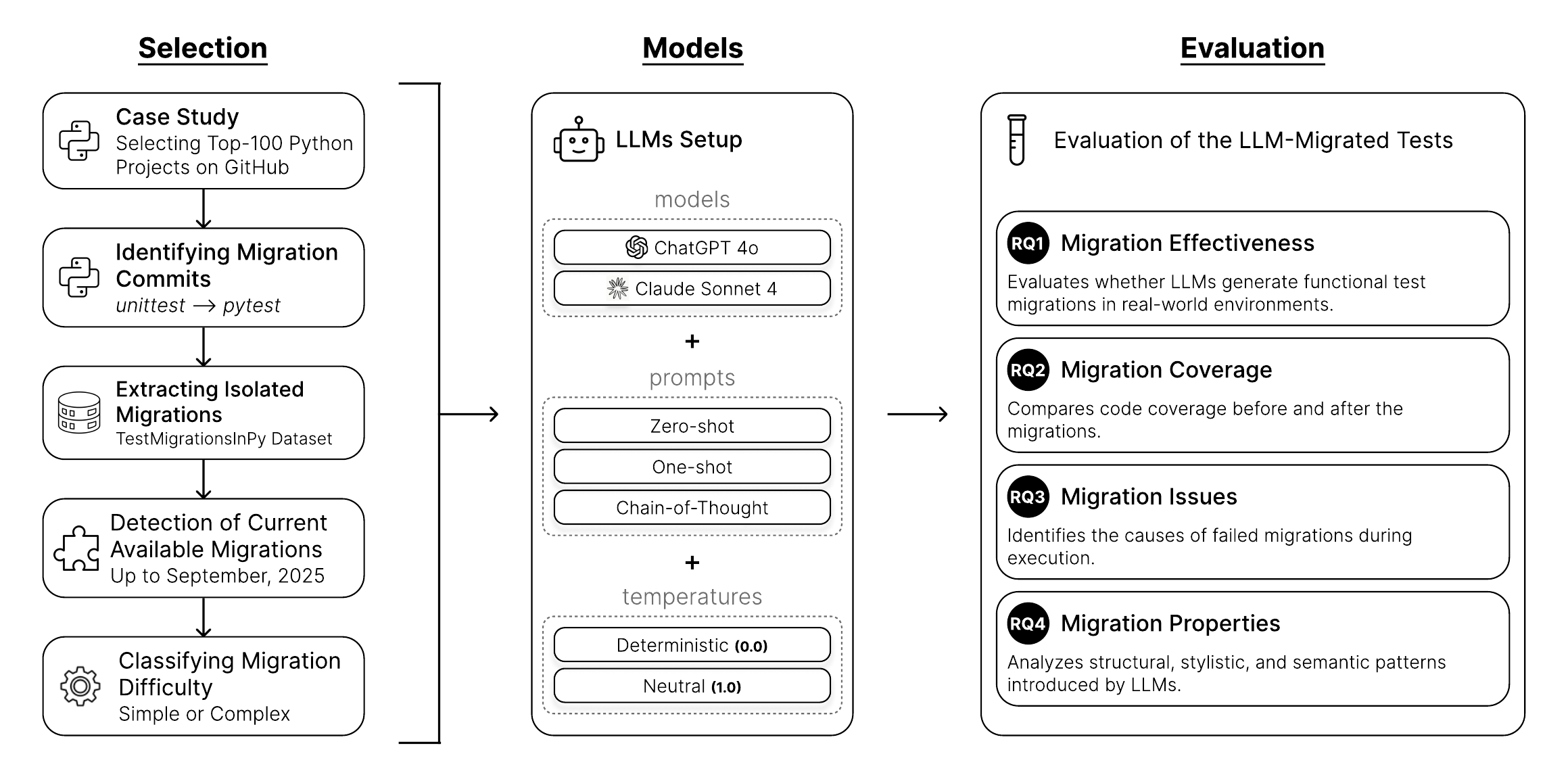}
  \caption{Overview of the study design.}
  \Description{Overview of the study design.}
  \label{fig:overview}
\end{figure*}

Figure~\ref{fig:overview} summarizes our study design. The methodology begins with selecting real-world Python projects and detecting framework migrations. Next, migration commits are analyzed to build the \textit{TestMigrationsInPy} dataset.
We then selected migrations that were still available in the latest commit of their projects.
These migrations were used in our experiment, where we asked LLMs to generate the migration for each scenario, replaced the original migration with the one produced by the LLM, and evaluated its correctness and coverage. 
Next, we detail each step.

\subsection{Case Study}

In this study, we analyze real-world and relevant software systems.
We analyze the top-100 Python projects with the most stars on GitHub.
This metric is widely adopted as a measure of the popularity of software projects~\cite{icsme2016, jss-2018-github-stars}.
These 100 projects came from a prior study that empirically analyzed the migration from unittest to Pytest~\cite{barbosa2022}.
The set includes projects broadly adopted worldwide, such as Pandas (data analysis library), Flask (web development framework), Requests (library for performing HTTP requests), and Ansible (open-source automation software).\footnote{The complete list of systems can be found in the original dataset: \url{https://doi.org/10.5281/zenodo.5847361}.}

\subsection{Detecting Migrations from Unittest to Pytest}

We relied on the tool proposed by Barbosa and Hora~\cite{barbosa2022} to detect projects that migrated from unittest to Pytest.
Basically, the tool traverses throughout the commit history and analyzes the \emph{removed} and \emph{added} lines of each commit.
One commit is considered \emph{migration commit} if at least one of the following rules applies:

\begin{enumerate}

  \item \textbf{Assert migration}: the commit removes unittest \mbox{\texttt{self.\-assert*}} and adds \mbox{\texttt{assert}} keyword.

  \item \textbf{Fixture migration}: the commit removes unittest fixtures (e.g., \mbox{\texttt{setUp}} and \mbox{\texttt{tearDown}}) and adds Pytest fixtures (e.g., \mbox{\texttt{@pytest.fixture}}).

  \item \textbf{Import migration}: the commit removes \mbox{\texttt{import unittest}} and adds \mbox{\texttt{import Pytest}}.

  \item \textbf{Skip migration}: the commit removes unittest test skips (e.g., \mbox{\texttt{@unittest.skipIf}}) and adds Pytest test skips (e.g., \mbox{\texttt{@Pytest.mark.skipif}}).

  \item \textbf{Expected failure migration}: the commit removes unittest expected failure (i.e., \mbox{\texttt{@unittest.expectedFailure}}) and adds Pytest expected failure (i.e., \mbox{\texttt{@Pytest.mark.xfail}}).
  
\end{enumerate}

For our study, we executed the migration detection tool on the top-100 selected projects, which detected 690 migration commits in 37 projects.
This is a significant increase when compared to the original study, which found 330 migration commits in 27 projects~\cite{barbosa2022}.

\subsection{Extracting Isolated Migrations}
\label{sec:isolated}

The next step in our study design is to create a dataset of migrations that can be used in our research as a ground truth.

We briefly describe the framework used to construct our \emph{TestMigrationsInPy} dataset~\cite{alves2025testmigrationsinpy}.\footnote{The real dataset name is omitted due to the double-blind review.}
The dataset was built based on the manual analysis of the migration commits collected in the previous step. 
It is important to notice that a migration commit may have one or more migrations from unittest to Pytest.
However, it is well-known that commits may include unrelated (i.e.,~tangled) changes~\cite{dias2015untangling}, e.g.,~it may perform migration and add/remove/update assertions.
To avoid this problem, we focused on detecting \emph{isolated migrations}, that is, migrations that simply replace unittest with Pytest, and no other unrelated changes are involved.
Moreover, to avoid noise caused by large commits, we filtered commits that modified more than 5 files.
Of the 690 migration commits collected in the previous step, we manually detected 923 isolated migrations that are used to create our dataset \emph{TestMigrationsInPy}.
This dataset is adopted as the ground truth in this research, but it can be used by any other research in the context of framework migration.


\begin{center}
\fcolorbox{black}{gray!15}{
  \parbox{\dimexpr1\linewidth-2\fboxsep-2\fboxrule}{
\textbf{Dataset:} \emph{TestMigrationsInPy} contains 923 real-world migrations from unittest to Pytest (\url{https://github.com/altinoalvesjunior/TestMigrationsInPy}).
  }
}
\end{center}


\subsection{Detection of Current Available Migrations}
\label{sec:selection}

We applied a multi-step filtering process to keep only migrations that can be validated, i.e.,~can be executed at the time this study is conducted.
From the 923 migrations available in the \textit{TestMigrationsInPy} dataset, we first checked whether the migrated test files were present in the latest project versions.
We then manually verified whether the migrated tests were still present in those files.
Finally, we reproduced the development environment of each project to ensure it could be built and executed locally.
For this paper, we only included for analysis the migrations satisfying all these criteria, considering project states up to September~30,~2025. 

After applying these filtering steps, 883 out of the 923 examined migrations were excluded. In most cases (53\%), the test files still existed, but the migrated methods were removed or their logic had changed substantially. Another 38\% showed partial preservation, with only some methods remaining, often refactored or simplified. The remaining 9\% referred to test files that had been deleted.

\begin{figure*}[t]
     \centering
         \includegraphics[width=0.95\textwidth]{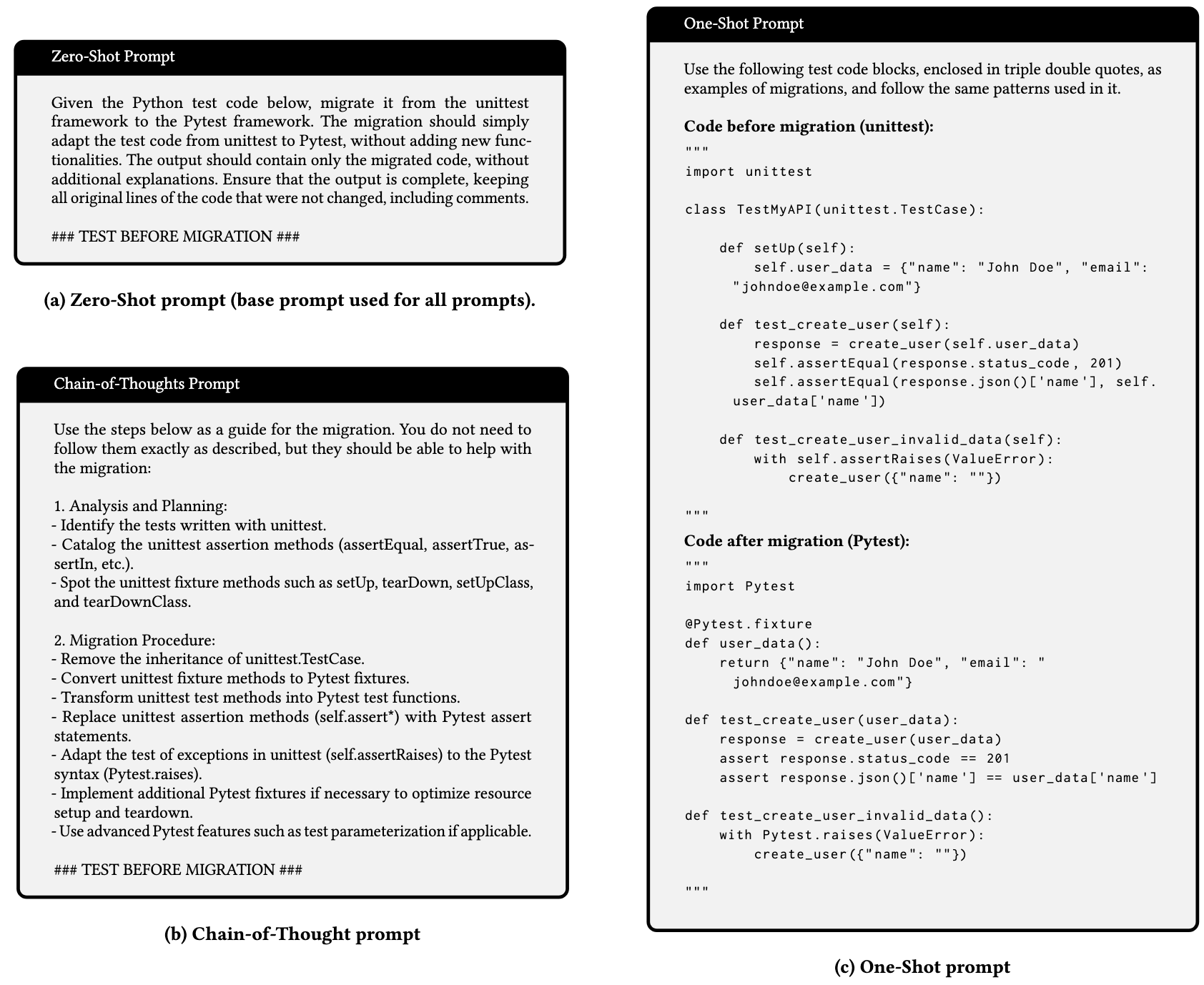}
         \caption{Prompt strategies adopted in this work.}
         \Description{Prompt strategies adopted in this work.}
        \label{fig:prompts}
\end{figure*}

Finally, we selected 40 migrations to assess in this research.
From these 40 migrations, 30 are classified as simple and 10 as complex cases according to our classification in Section~\ref{sec:migration-difficult}. 
These 40 migrations originated from seven projects, as presented in Table~\ref{tab:projects}.

\begin{table}[h]
\centering
\footnotesize
\setlength{\tabcolsep}{6pt}
\renewcommand{\arraystretch}{1.15}
\caption{Origin of the selected migrations.}
\label{tab:projects}
\begin{tabular}{@{}lcl@{}}
\toprule
\textbf{owner/name} & \textbf{Stars} & \textbf{URL} \\ 
\midrule
\textit{ansible/ansible} & 66{,}900 & \url{https://github.com/ansible/ansible} \\
\textit{apache/airflow} & 43{,}000 & \url{https://github.com/apache/airflow} \\
\textit{ray-project/ray} & 39{,}600 & \url{https://github.com/ray-project/ray} \\
\textit{httpie/httpie} & 36{,}900 & \url{https://github.com/httpie/httpie} \\
\textit{cookiecutter/cookiecutter} & 22{,}200 & \url{https://github.com/cookiecutter/cookiecutter} \\
\textit{beetbox/beets} & 14{,}100 & \url{https://github.com/beetbox/beets} \\
\textit{redis/redis-py} & 13{,}300 & \url{https://github.com/redis/redis-py} \\
\bottomrule
\end{tabular}
\end{table}

\subsection{Classifying Migration Difficulty}
\label{sec:migration-difficult}

As discussed in Section~\ref{sec:background}, not all migrations have the same level of difficulty. For example, migrations involving only changes of assertions are potentially easier to perform than migrations of fixtures.
To further explore this variation, we classify the migration into \emph{simple} or \emph{complex}.

\emph{Simple migrations} involve only direct changes from unittest to Pytest, such as converting classes that inherit from \texttt{unittest.\-Test\-Case} into standalone test functions and replacing unittest assertions (e.g., \texttt{self.assertEqual(a,b)}) with native assert statements.
An example of a simple migration is \texttt{test\_candle\_int\_4} from the termgraph project.\footnote{\url{https://github.com/sgeisler/termgraph/commit/d5665248b7d596cabe0a5}}
\emph{Complex migrations}, on the other hand, require structural adaptations and the use of additional Pytest resources. They typically involve fixtures, refactoring of \texttt{setUp}/\texttt{tearDown}, mocks, asynchronous tests, or interactions with external components such as I/O and/or network resources. They may also include adjustments to assertion logic.
An example of a complex migration is \texttt{setup\_attrs} from the airflow project.\footnote{\url{https://github.com/prompt-toolkit/pyvim/commit/7e1c7bfb505cefba468}}

\subsection{LLMs Setup}

We selected OpenAI’s GPT 4o because it represents the state of the art in code generation~\cite{openai2023gpt4}.
We also selected Anthropic’s Claude Sonnet 4 due to its consistent superiority over other large language models---including OpenAI’s, Google’s Gemini, and Deep\-Seek---in activities involving code understanding and generation~\cite{gao2025comparison, app15189907}.

The prompt used to evaluate both models was inspired by prior research on library migration~\cite{esem2024_api_migration_llm}.
Specifically, we explore three different prompting methods: Zero-Shot~\cite{sahoo2024systematic}, One-Shot~\cite{chen2025unleashing} and Chain Of Thought~\cite{wei2022chain}.
While the Zero-Shot approach contains only the task description, the One-Shot strategy adds a single example, allowing the model to have access to a desired structure before generating the answer.
The Chain-of-Thought approach introduces a step-by-step reasoning process that leads the model toward the outcome.
Figure \ref{fig:prompts} details the prompts used in our research.

To ensure a balanced evaluation between determinism and creativity, we considered two temperature values: 0.0 (deterministic) and 1.0 (neutral).
The deterministic value produces reproducible outputs, reflecting the model’s reliability in constrained scenarios~\cite{renze2024effect}.
The neutral configuration preserves the model’s default probability distribution and enables a balanced degree of creativity and diversity in the generated code~\cite{li2025exploring}.

Finally, for each of the 40 selected migrations, we created three prompts (Zero-Shot, One-Shot, and Chain-of-Thoughts), and executed each twice (under deterministic and neutral configurations) for both models (GPT 4o and Claude Sonnet 4).
As a result, we performed 480 requests, i.e.,~40 migrations $\times$ 3 prompts $\times$ 2 temperatures $\times$ 2 models.
For this purpose, we developed a script to access GPT 4o and Claude Sonnet 4 APIs.
The results were stored in structured format containing the migration returned as response and the request setup,i.e.,~model, prompt strategy, and temperature.

\subsection{Evaluation of the LLM-Migrated Tests}
\label{sec:evaluation}

To evaluate the effectiveness of the migration process, we executed all the \emph{LLM-migrated test} on the original projects' test suites.
For each project, we cloned the latest available tag or release on GitHub (up to September 30th, 2025) into a local environment to reproduce its original setup. 
For example, we installed all required dependencies, downloaded necessary Docker images, and configured auxiliary components.
The goal was to \emph{ensure} a fully functional testing environment.

Next, we manually edited the original test files and replaced the migrated source code with one of the LLM-migrated versions; the rest of the project remained unchanged.
For each replacement, we executed the test suite and recorded whether the test case had \emph{PASSED} or \emph{FAILED}. We also collected the project-level coverage for every execution using \texttt{coverage.py}~\cite{coverage-py}.
This allowed us to quantify the structural impact of each migration and compare it against the baseline coverage of the original tests.

We examined the differences between the migration performed by the LLMs and the original one performed by the developers.
Specifically, we first looked for duplicated answers to quantify how often the LLMs generated equivalent code despite variations in prompting configuration.
Next, we extracted structural elements important to implement tests in Pytest, such as \texttt{assert}, \texttt{@pytest.\-fixture}, \texttt{pytest.raises}.

\subsection{Research Questions}


The goal of this study is to evaluate the effectiveness of LLMs at migrating tests from \texttt{unittest} to \texttt{Pytest}.
We assessed the migrations generated by the LLMs with respect to their correctness, coverage, issues, and changes.
We propose four research questions:

\noindentparagraph{\textbf{RQ1: Can LLMs generate correct test migrations?}}
This question explores whether the LLMs generate valid migrations.
We consider that LLM-generated migrations are correct when they can be successfully executed in the test environment of their projects.
We explore the migration in five perspectives: (i) model, (ii) temperature, (iii) strategy, (iv) difficulty, and (v) project.

\noindentparagraph{\textbf{RQ2: Do LLMs-generated migrations change test coverage?}} 
This question evaluates whether LLM-generated migrations not only produce valid tests, but if they keep their original intent.
We hypothesize that changes in code coverage mean that a correct but different test was generated by the LLM.

\noindentparagraph{\textbf{RQ3: What are the errors in LLMs-generated migrations tests?}} 
Here, we investigate the cases where LLM-generated migrations failed to execute. 
This analysis aims to understand the underlying causes of these failures, such as structural inconsistencies, missing methods, or logical inconsistencies introduced during migration.

\noindentparagraph{\textbf{RQ4: Which changes are mostly introduced in LLMs-generated migrations tests?}}
Finally, this RQ analyzes multiple properties of LLM-generated migrations, including changes in code structure and duplicated migrations.

\section{Results}


\begin{table*}[!t]
\centering
\footnotesize
\setlength{\tabcolsep}{8pt}
\caption{Successful migrations by model, temperature, strategy, and difficulty.}
\label{tab:effectiveness}
\begin{tabular*}{\textwidth}{@{\extracolsep{\fill}} l c l cccc c}
\toprule
\multirow{3}{*}{\textbf{Migration Difficulty}} & \multirow{3}{*}{ \textbf{\#Migrations}} & \multirow{3}{*}{\textbf{Prompt Strategy}} & \multicolumn{4}{c}{\textbf{Model}} & \multirow{3}{*}{\textbf{Total}} \\
\cmidrule(lr){4-7}
& & & \multicolumn{2}{c}{\textbf{GPT 4o}} & \multicolumn{2}{c}{\textbf{Claude Sonnet 4}} & \\
\cmidrule(lr){4-5}\cmidrule(l){6-7}
& & & Deterministic (0.0) & Neutral (1.0) & Deterministic (0.0) & Neutral (1.0) & \\
\midrule
\multirow{3}{*}{\makecell[l]{\textbf{Simple}\\\emph{(assertions and raises)}}}
&  & Zero-shot         & 15 & 15 & 16 & 16 & 62 \\
& 30 & One-shot          & 14 & 15 & 15 & 15 & 59 \\
&  & Chain-of-Thought & 16 & 16 & 16 & 16 & 64 \\
\addlinespace[2pt]
\midrule
\multirow{3}{*}{\makecell[l]{\textbf{Complex}\\\emph{(fixtures, mocks, integrations, I/O)}}}
&  & Zero-shot         & 2 & 2 & 6 & 6 & 16 \\
& 10 & One-shot          & 4 & 3 & 4 & 4 & 15 \\
&  & Chain-of-Thought & 3 & 4 & 5 & 5 & 17 \\
\midrule
\textbf{Total} & & & 54 & 55 & 62 & 62 & 233 \\
\bottomrule
\end{tabular*}
\end{table*}

\subsection{RQ1: Correctness of Test Migrations}
\label{sec:effectiveness}




We find that 247 out of 480 (51.5\%) LLM-generated migrations have failed, i.e.,~returned \textit{FAILED}.
Conversely, 233 migrations were successfully executed, i.e.,~returned \textit{PASSED}.
Although the number of successful migrations almost matches the unsuccessful ones, most migrations did not preserve the functional behavior of the original tests.
The remainder of this section focuses on analyzing the successful migrations in five perspectives: (i) model, (ii) temperature, (iii) strategy, (iv) difficulty, and (v) project.


\subsubsection{\textbf{By Model}}

Table~\ref{tab:effectiveness} shows that Claude Sonnet 4 achieved slightly higher correctness, with 124 (51.66\%) migrations returned \textit{PASSED}, while GPT-4o reported 109 (45.41\%).
The difference of 6.25 points between the models is modest, but suggests that Claude may handle certain contextual and semantic aspects of the migration process more reliably. 
This benefit might be due to differences in how the models are built and fine-tuned. 
Recent research shows how these differences can affect the way LLMs find errors and change code~\cite{boukhlif2024llms, ramler2025}.


\subsubsection{\textbf{By Temperature}} 
Claude Sonnet 4 maintains the same correctness at neutral and deterministic approaches, with 62 migrations \textit{PASSED} in each case (124 in total, 53.21\%), indicating low sensitivity to temperature, as shown in Table~\ref{tab:effectiveness}. 
GPT-4o shows only a minimum increase from 54 (23.17\%) in \emph{deterministic} to 55 (23.60\%) in \emph{neutral}. 
These results suggest that temperature has a negligible impact on correctness.
Prior studies report similar stability across temperature ranges in reasoning and code-generation tasks~\cite{renze2024temperature, arora2024}. 


\subsubsection{\textbf{By Strategy}}

Overall, the Chain-of-Thought strategy yielded the best overall correctness, with 81 migrations marked as \textit{PASSED} (34.76\%) across both models, as summarized in Table~\ref{tab:effectiveness}.
Claude Sonnet 4 performed better under the Zero-shot approach, with 44 migrations returned \textit{PASSED} (35.48\%) out of the model’s 124 successful executions.
GPT-4o, on the other hand, achieved its best correctness with the Chain-of-Thought strategy, with 39 (35.78\%) migrations successfully executed out of the model’s 109 successful cases.
Conversely, the One-shot approach yielded lower success rates for both models.
Prior studies have shown that example-based prompting can induce representational bias or over-specialization toward the provided instance, limiting output diversity and generalization \cite{bender2021dangers, 10.1145/3672608.3707812}. 
A single example may lead LLMs to anchor their results on that specific instance and generate migrations consistent with the example but misaligned with other valid cases. 


\subsubsection{\textbf{By Difficulty}} 

\textit{Simple} migrations presented higher success rates in both models, confirming that tasks requiring just syntactic adjustments are more easily captured by LLMs, as shown in Table~\ref{tab:effectiveness}. 
The best results were achieved under the Chain-of-Thought strategy, with 64 (27.46\%) migrations performed successfully for both models. 
In contrast, the results vary more in \textit{Complex} migrations.
While Claude Sonnet was able to provide proper answers for 44 of the migration scenarios (Zero-shot), GPT-4o achieved its best (One-Shot and Chain-of-Thought).
Nevertheless, both models struggled to preserve test logic structures such as fixtures and mocks.

\subsubsection{\textbf{By Project}} 

Lastly, Table~\ref{tab:project-summary} summarizes the results by project.
We detail the number of analyzed migrations, the distribution between simple and complex migrations, the total number of test executions--based on the 480 migration requests---and the percentage of successful outcomes.

The \textit{ansible/ansible} project achieved the highest effectiveness, with 168 (99.4\%) of successful executions; all successful migrations were classified as simple. 
Interestingly, \textit{redis/redis-py} had a similar number of simple migrations, but none of them executed successfully.
The \textit{httpie/cli} project obtained the second-best performance (156, 83.3\%).
Finally, \textit{apache/airflow} and \textit{ray-project/ray} reported no successful executions.

\begin{table}[H]
\centering
\scriptsize
\caption{Successful migrations by project.}
\label{tab:project-summary}
\setlength{\tabcolsep}{6pt}
\begin{tabular}{@{}lccccccc}
\toprule
\multirow{2}{*}{\textbf{Project}} &
\multicolumn{3}{c}{\textbf{Number of Migrations}} &
    \multirow{2}{*}{\textbf{Exec.}} &
\multirow{2}{*}{\textbf{Success}} & \multirow{2}{*}{\textbf{\%}} \\
\cmidrule(lr){2-4}
& \textbf{\#} & \textbf{Simple} & \textbf{Complex} & & \\
\midrule
ansible/ansible         & 14  & 13  & 1  & 168 & 167 & 99.4\% \\
httpie/httpie           & 2  & 1  & 1  & 24  & 20 & 83.3\% \\
beetbox/beets           & 4  & 1  & 3  & 48  & 30 & 62.5\%  \\
cookiecutter/cookiecutter & 3  & 1  & 2  & 36  & 16 & 44.4\%  \\
redis/redis-py          & 13 & 13 & 0  & 156 & 0 & 0\%   \\
apache/airflow          & 3  & 0  & 3  & 36  & 0 & 0\%   \\
ray-project/ray         & 1  & 1  & 0  & 12  & 0 & 0\%   \\
\midrule
\textbf{Total}          & 40 & 30 & 10  & 480 & 233 & -\\
\bottomrule
\end{tabular}
\end{table}

\begin{table*}[!t]
\centering
\footnotesize
\setlength{\tabcolsep}{4pt}
\renewcommand{\arraystretch}{1.15}
\caption{Unsuccessful migrations by model, temperature, strategy, and difficulty.}
\label{tab:error-types}
\begin{tabular*}{\textwidth}{@{\extracolsep{\fill}} ll *{12}{c} c}
\toprule
\multirow{3}{*}{\textbf{Error}} & \multirow{3}{*}{\textbf{owner/project}} &
\multicolumn{6}{c}{\textbf{GPT 4o}} &
\multicolumn{6}{c}{\textbf{Claude Sonnet 4}} &
\multirow{3}{*}{\textbf{Total}} \\
\cmidrule(lr){3-8}\cmidrule(lr){9-14}
& & 
\multicolumn{2}{c}{\textbf{Zero-shot}} & 
\multicolumn{2}{c}{\textbf{One-shot}} & 
\multicolumn{2}{c}{\textbf{Chain-of-Thought}} &
\multicolumn{2}{c}{\textbf{Zero-shot}} & 
\multicolumn{2}{c}{\textbf{One-shot}} & 
\multicolumn{2}{c}{\textbf{Chain-of-Thought}} & \\
\cmidrule(lr){3-4}\cmidrule(lr){5-6}\cmidrule(lr){7-8}
\cmidrule(lr){9-10}\cmidrule(lr){11-12}\cmidrule(lr){13-14}
& & 
\textbf{0.0} & \textbf{1.0} & 
\textbf{0.0} & \textbf{1.0} & 
\textbf{0.0} & \textbf{1.0} &
\textbf{0.0} & \textbf{1.0} & 
\textbf{0.0} & \textbf{1.0} & 
\textbf{0.0} & \textbf{1.0} & \\
\midrule
AssertionError & \textit{ray-project/ray} & 1 & 1 & 1 & 1 & 1 & 1 & 1 & 1 & 1 & 1 & 1 & 1 & 12 \\
Missing Fixtures & \textit{httpie/httpie} & 0 & 0 & 1 & 1 & 0 & 0 & 0 & 0 & 1 & 1 & 0 & 0 & 4 \\
Signature Drift & \textit{beetbox/beets} & 3 & 3 & 2 & 3 & 3 & 2 & 0 & 0 & 1 & 1 & 0 & 0 & 18 \\
Structural Mismatch & \textit{cookiecutter/cookiecutter} & 3 & 3 & 1 & 1 & 1 & 1 & 1 & 1 & 2 & 2 & 2 & 2 & 20 \\
SyntaxError & \textit{ansible/ansible} & 0 & 0 & 1 & 0 & 0 & 0 & 0 & 0 & 0 & 0 & 0 & 0 & 1 \\
 & \textit{apache/airflow}, & 3 & 3 & 3 & 3 & 3 & 3 & 3 & 3 & 3 & 3 & 3 & 3 & 36 \\
TypeError & \textit{redis/redis-py} & 13 & 13 & 13 & 13 & 13 & 13 & 13 & 13 & 13 & 13 & 13 & 13 & 156 \\
\midrule
\textbf{Total} &  & 23 & 23 & 22 & 22 & 21 & 20 & 18 & 18 & 21 & 21 & 19 & 19 & 247 \\
\bottomrule
\end{tabular*}
\end{table*}

\begin{center}
\fcolorbox{black}{gray!15}{
  \parbox{\dimexpr1\linewidth-2\fboxsep-2\fboxrule}{
\textbf{RQ1.} Overall, 48.5\% (233 out of 480) of all migrations executed successfully.
Chain-of-Thought yielded the best overall correctness.
Claude Sonnet 4 achieved slightly higher correctness than GPT-4o.
\textit{Simple} migrations presented higher success rates in both models, but Claude Sonnet performed better in \textit{complex} ones.
  }
}
\end{center}


\subsection{RQ2: Coverage in Test Migrations}

In this RQ, we explore whether LLM-generated tests change the test coverage as compared to the original migration.
Changes in test coverage may indicate that a correct but different test was generated by the models.

We find that the coverage remained exactly the same for all successful cases.
For instance, \textit{ansible/ansible} maintained 39\%, \textit{cookiecutter/cookiecutter} 33\%, \textit{httpie/httpie} 94\%, and \textit{beetbox/beets} 62.96\%. 
These results suggest that LLM-generated migrations effectively reproduced the test behavior.

\begin{center}
\fcolorbox{black}{gray!15}{
  \parbox{\dimexpr1\linewidth-2\fboxsep-2\fboxrule}{
\textbf{RQ2.} Test coverage remained unchanged in successful migrations, i.e.,~the LLM-generated migrations preserved the test behavior.
  }
}
\end{center}


\subsection{RQ3: Errors in Test Migrations}

This RQ explores the 247 cases where LLM-generated migrations failed to execute.
We identified six major categories of errors across such LLM-generated test migrations that failed to execute.
These categories are described below and represent distinct failure patterns observed during the test execution:



\begin{itemize}

\item \textbf{AssertionError}: subtle changes that led to failed assertions, such as rounding errors, string formatting changes, or different expected values.

\item \textbf{Structural Mismatch}: inconsistencies in the structural style or test organization that caused failures in test discovery or disrupted class-based dependencies.

\item \textbf{Missing Fixtures}: missing references to fixtures, clients, or variables used in the tests.

\item \textbf{TypeError}: misuse of data types or interactions with uninitialized objects, frequently linked to absent mocks or clients.

\item \textbf{Signature Drift}: undesired modifications in function signatures, such as added, removed, or renamed parameters that cause incompatibility with other test components.

\item \textbf{SyntaxError (Parse)}: malformed instructions in the code, such as unbalanced parentheses, malformed \texttt{try/except} blocks, or indentation issues.
\end{itemize}

Table~\ref{tab:error-types} details the results per category.
Across all executions, a total of 247 failures were recorded out of 480 migrations, encompassing all combinations of models, prompting strategies, and temperature settings. 
Among the 40 unique migrations, 25 experienced at least one failure. 
When analyzing failures across configurations, we observed that GPT-4o produced the highest number of failing migrations under the Zero-shot strategy, while Claude Sonnet 4 did so under the One-shot strategy. 
Despite these variations, temperature had virtually no influence on the outcome—error counts remained identical between \textit{Deterministic (0.0)} and \textit{Neutral (1.0)} temperatures for all strategies, except for a marginal difference of one case in the Chain-of-Thought configuration for GPT-4o.

Interestingly, the number of errors reported by some types remained the same regardless of the LLM setup.
For example, the number of \textit{TypeError} remained 13 between all configurations; similar behavior happens with \textit{SyntaxError} detected in \textit{apache/airflow} and \textit{AssertionError}.
In \textit{redis/redis-py}, most errors were caused by missing dependencies and incorrect fixture substitutions, which led to \texttt{TypeError} exceptions. Similarly, \textit{apache/airflow} frequently failed due to malformed control blocks (e.g.,~unbalanced \texttt{try/except} statements), resulting in \texttt{SyntaxError} during module imports.

In contrast, \textit{ansible/ansible} and \textit{httpie/httpie} have more localized failures, mainly caused by broken test discovery when migrated functions were moved out of their original classes or lost their fixture associations. Finally, in \textit{beetbox/beets} and \textit{cookiecutter/cookiecutter}, signature drift and inheritance modifications disrupted shared setup routines, producing integration inconsistencies that propagated across dependent tests. 


\begin{center}
\fcolorbox{black}{gray!15}{
  \parbox{\dimexpr1\linewidth-2\fboxsep-2\fboxrule}{
    \textbf{RQ3.} 
    Overall, 51.5\% (247 out of 480) of all migrations failed the execution.
    They represent 25 out of the 40 unique migrations. 
    Most issues involved problems in handling dependencies, incorrect fixture usage, or inconsistencies in structural elements, such as test setup and inheritance. 
  }
}
\end{center}


\subsection{RQ4: Changes in Test Migrations}
\label{sec:rq4}

To guide this analysis, we organized the discussion into two perspectives: (i) changes in code structure and (ii) duplicated migrations.

\subsubsection{\textbf{Changes in Code Structure}}

LLMs exhibited distinct structural changes when migrating tests from \texttt{unittest} to \texttt{Pytest}, as detailed in Table~\ref{tab:model-prompt-temp}.
Overall, Claude Sonnet 4 presented a conservative migration style, preserving most of the original test code across the multiple configurations.
For example, Claude preserved class-oriented organization in 70.83\% of migrations and maintained references to \texttt{unittest} in 33.75\%. 
In contrast, GPT-4o performed more substantial changes.
Classes were preserved in 45\% of the cases, the \texttt{unittest} package is mentioned in 9.17\% of the tests. 
Despite these differences, both models fully removed unitttest \texttt{self.assert*}, mostly adopting native Pytest \texttt{assert}.

Related to prompt strategies, we observe that the number of classes declined substantially (Claude 49\%, GPT 2.5\%) when using One-shot; the number of \texttt{import pytest} statements also increased in this scenario.
The Zero-shot, on the other hand, generated outputs structured and aligned with the original tests. 
Claude preserved class structures in 100\% of cases and GPT in 88\%.
Pytest's advanced commands are more present in migrations generated by the \textit{Chain-of-Thougths} strategy.
Both GPT-4o and Claude achieved the highest results for \texttt{@pytest.fixture} and \texttt{pytest.raises()} in migrations performed using this prompt.

Temperature variation had a negligible impact on these trends.
Across all prompting settings, the difference between deterministic and neutral configurations remained below 2.5 p.p. for most metrics. Nonetheless, the deterministic approach produced more consistent code---presented lower variations in fixture usage and import organization---whereas neutral generations introduced only minor stylistic differences without semantic impact.

\begin{table*}[!t]
\centering
\scriptsize
\setlength{\tabcolsep}{3.5pt}
\caption{Code metrics \textit{before} and \textit{after} migration across models, prompts, and temperatures.}
\label{tab:model-prompt-temp}
\resizebox{\textwidth}{!}{%
\begin{tabular}{@{}l l c c c c c c c c c c c@{}}
\toprule
\textbf{Model} & \textbf{Prompt} & \textbf{Temp.} & \textbf{Total} & \textbf{\textit{class}} & \textbf{\textit{test\_*}} & \makecell{\textbf{\textit{import}}\\\textbf{\textit{Pytest}}} & \makecell{\textbf{\textit{import}}\\\textbf{\textit{unittest}}} & \textbf{\textit{@pytest.fixture}} & \textbf{\textit{pytest.raises()}} & \textbf{\textit{self.assert*}} & \makecell{\textbf{\textit{assert}}\\\textbf{(Pytest)}} & \makecell{\textbf{\textit{LOC}}\\\textbf{(mean)}} \\
\midrule
Code (before) & -- &  & 40 & 100\% & 87.50\% & 7.50\% & 100\% & 0\% & 0\% & 85\% & 10\% & 25 \\
\midrule
\multirow{8}{*}{\makecell[l]{GPT 4o}}
  & Zero-shot & 1.0 & 40 & 87.50\% & 90\% & 40\% & 10\% & 5\% & 15\% & 0\% & 77.50\% & 19 \\
  &           & 0.0 & 40 & 87.50\% & 90\% & 40\% & 7.50\%  & 5\% & 15\% & 0\% & 85\% & 19 \\
  \cmidrule(lr){2-13}
  & One-shot  & 1.0 & 40 & 2.50\%  & 92.50\% & 100\% & 7.50\%  & 22.50\% & 15\% & 0\% & 85\% & 19 \\
  &           & 0.0 & 40 & 2.50\%  & 90\%  & 97.50\%  & 7.50\%  & 22.50\% & 15\% & 0\% & 85\% & 19 \\
  \cmidrule(lr){2-13}
  & CoT       & 1.0 & 40 & 42.50\% & 92,50\% & 87.50\%  & 12,50\% & 32.50\% & 17.50\% & 0\% & 87.50\% & 20 \\
  &           & 0.0 & 40 & 47.50\% & 90\% & 90\%  & 10\% & 30\% & 17.50\% & 0\% & 85\% & 20 \\
  \cmidrule(lr){2-13}
  & \textit{Total} &   & 240 & 45\% & 90.83\% & 75.83\% & 9.17\% & 19.58\% & 15.83\% & 0\% & 84.17\% & 19 \\
\midrule
\multirow{8}{*}{\makecell[l]{Claude\\Sonnet 4}}
  & Zero-shot & 1.0 & 40 & 100\% & 90\% & 40\% & 20\% & 0\% & 15\% & 0\% & 85\% & 19 \\
  &           & 0.0 & 40 & 100\% & 90\% & 42.50\% & 17.50\% & 0\% & 15\% & 0\% & 85\% & 19 \\
  \cmidrule(lr){2-13}
  & One-shot  & 1.0 & 40 & 50\%  & 90\% & 95\% & 45\% & 7.50\% & 15\% & 0\% & 85\% & 20 \\
  &           & 0.0 & 40 & 47.50\%  & 90\% & 92.50\% & 35\% & 7.50\% & 15\% & 0\% & 85\% & 20 \\
  \cmidrule(lr){2-13}
  & CoT       & 1.0 & 40 & 65\%  & 90\% & 80\% & 42.50\% & 7.50\% & 15\% & 0\% & 85\% & 19 \\
  &           & 0.0 & 40 & 62.50\%  & 90\% & 82.50\% & 42.50\% & 12.50\% & 17.50\% & 0\% & 85\% & 19 \\
  \cmidrule(lr){2-13}
  & \textit{Total} &   & 240 & 70.83\% & 90\% & 72.08\% & 33.75\% & 5.83\% & 15.42\% & 0\% & 85\% & 19 \\
\bottomrule
\end{tabular}%
}
\end{table*}

\subsubsection{\textbf{Duplicated Migrations}}

Overall, we identified 117 duplicated migration groups, encompassing 348 variants (72.5\% of all generated migrations). Each group represents a set of identical migration outputs produced under different model or configuration settings, while variants refer to the individual instances within those groups. This highlights a marked convergence in model behavior, where different configurations frequently result in identical code generations.
As shown in Table~\ref{tab:duplication}, Claude Sonnet 4 exhibited the highest degree of redundancy, with 89.17\% of its variants being identical across at least one configuration, compared to 55.83\% for GPT-4o. At the prompt level, Claude maintained near-deterministic behavior, reaching duplication rates of 92.50\% under Zero-shot, 90\% under One-shot, and 85\% under Chain-of-Thought. In contrast, GPT produced more diverse and variable generations, with duplication ranging from 67.50\% in Zero-shot to 45\% in Chain-of-Thought.

\begin{table}[H]
\centering
\small
\setlength{\tabcolsep}{4pt}
\caption{Duplicated migrations.}
\label{tab:duplication}
\begin{tabular}{@{}l l c c c@{}}
\toprule
\textbf{Model} & \textbf{Prompt} & \textbf{Temp.} & \textbf{Total} & \textbf{Duplications} \\
\midrule
\multirow{6}{*}{\textbf{GPT 4o}} 
  & Zero-Shot         & 0.0 & 40 & 26 \\
  &                   & 1.0 & 40 & 27 \\
  \cmidrule(lr){2-5}
  & One-Shot          & 0.0 & 40 & 23 \\
  &                   & 1.0 & 40 & 22 \\
  \cmidrule(lr){2-5}
  & Chain-of-Thought & 0.0 & 40 & 18 \\
  &                   & 1.0 & 40 & 18 \\
\midrule
\multirow{6}{*}{\textbf{Claude Sonnet 4}} 
  & Zero-Shot         & 0.0 & 40 & 37 \\
  &                   & 1.0 & 40 & 37 \\
  \cmidrule(lr){2-5}
  & One-Shot          & 0.0 & 40 & 36 \\
  &                   & 1.0 & 40 & 36 \\
  \cmidrule(lr){2-5}
  & Chain-of-Thought & 0.0 & 40 & 34 \\
  &                   & 1.0 & 40 & 34 \\
\midrule
\textbf{Total} &  &  & \textbf{480} & \textbf{328} \\
\bottomrule
\end{tabular}
\end{table}

Temperature variation had minimal impact on duplication rates. Claude’s outputs remained nearly identical across the \textit{precise} (88.33\%) and neutral (90\%) configurations, while GPT consistently produced around 55.83\% duplicates in both. These observations suggest that prompt design—rather than temperature—plays the dominant role in driving convergence among generated code variants.

\begin{center}
\fcolorbox{black}{gray!15}{
  \parbox{\dimexpr1\linewidth-2\fboxsep-2\fboxrule}{
    \textbf{RQ4.} Claude Sonnet 4 preserved class-based structures and \texttt{unittest} elements, while GPT-4o favored function-based rewrites with \texttt{Pytest} elements.  
    Duplicated migrations were frequent, revealing convergence toward stable migration templates, especially for Claude (up to 92.50\%).
  }
}
\end{center}


\section{Discussion and Implications}

\subsubsection*{\textbf{LLMs underperformed the migration of complex cases.}} 
Both models achieved consistent results in simple migrations, but their performance declined in complex scenarios. 
While 51\% of simple migrations were executed successfully, this performance drops to 40\% when considering only complex ones (RQ1).
We believe that these failures are primarily due to the models’ limited capacity to analyze the architectural aspects of the tests being migrated. 
For example, most execution failures involved incorrect fixtures and test setup.
This behavior was also noticed in other works, where LLMs struggle to find the correct outcome in larger scenarios~\cite{silva2024detecting}.




\subsubsection*{\textbf{Temperature setup did not impact the migrations.}} 
Despite being theoretically designed to influence creativity and exploration, temperature had minimal impact on the migration outcomes. 
As observed in RQ1, both models produced almost identical structural outcomes across all configurations, i.e.,~they produced the same set of successful tests independent of migration difficulty or prompt strategy.
Similar behavior is also noted for failed tests (RQ3); a higher temperature value did not help the LLMs to overcome with an alternative solution that was correct.
These findings emphasize that, for local code transformation tasks, prompt design and access to relevant project context have more influence on outcome quality than random variation in model outputs~\cite{dong2025rethinking, du2024evaluating}.

\subsubsection*{\textbf{No prompt configuration outperformed the others.}} 
Across all prompt strategies, no configuration consistently outperformed the others. 
As observed in RQ1, while Chain-of-Thought occasionally produced slightly more structured migrations, the differences were minimal.
Despite the variations in prompting strategy, both models frequently generated correct migrations for similar scenarios, indicating a strong dependency on common transformation patterns. 
Once a migration pattern is explicitly mentioned---such as converting \texttt{setUp()} to \texttt{@pytest.fixture}---the LLMs started using it with more frequency~\cite{dong2025rethinking, chen2025memorization}.
By contrast, adding a source code example can occasionally introduce some bias to the migration process.
This happens with the One-shot prompt configuration, where it consistently added the same fixture-based setup pattern from the provided example---particularly inserting unnecessary \texttt{@pytest.fixture} functions even in tests that did not originally rely on shared fixtures.
This suggests that source code examples can constrain the LLMs capacity to provide a more diverse output. 


\subsubsection*{\textbf{Migration executions failed mostly due to context-sensitive instructions, such as fixtures and test setup.}}

Most of the execution failures observed in RQ3 originated from missing dependencies that were not explicitly represented in the test files. 
While the LLMs successfully translated the syntax from \texttt{unittest} to \texttt{Pytest}, they frequently ignored the relationship between fixtures, setup routines, and external components to the tests under migration. 
For instance, in \textit{redis/redis-py}, fixture-based clients were replaced with direct object instantiations (e.g., \texttt{redis.StrictRedis()}), breaking the controlled initialization of shared Redis connections and resulting in \texttt{TypeError} exceptions. 
Other failures, however, stemmed from structural and semantic inconsistencies rather than missing dependencies: in \textit{apache/airflow}, malformed control structures—such as unbalanced \texttt{try/except} blocks—caused syntax errors during import, while in \textit{ray-project/ray}, subtle rounding precision mismatches produced false \texttt{AssertionError} failures.
In either case, we found no LLM setup able to fix these changes, suggesting additional information is needed to properly detect and fix these errors.
These findings highlight the need for approaches that incorporate detailed, fine-grained information about the test environment, allowing LLMs to understand and leverage contextual relationships among test components.

\subsubsection*{\textbf{Advanced commands are used only if explicitly mentioned in the prompt.}}

As shown in RQ4, both models rarely applied advanced \texttt{Pytest} features unless explicitly mentioned in the prompt. 
Even though \texttt{Pytest} supports richer constructs, such as parameterization, for example, the models typically performed direct one-to-one transformations from \texttt{unittest}, avoiding higher-level abstractions. 
In other words, the models favor a conservative transformation over a more idiomatic one, even when these could lead to more maintainable test suites. 
This behavior is consistent with previous studies showing that LLMs focus on reproducing familiar syntax instead of fully understanding and applying the deeper semantics of the target framework~\cite{dong2025rethinking, du2024evaluating}. 
The effective automation of testing framework migrations passes by adopting techniques to encourage the model to generate more idiomatic outcomes, such as guided prompts and finetuning strategies~\cite{Improta2025, schafer2023empirical}.

\section{Threats to Validity}


\subsubsection*{\textbf{Internal Validity}}
Although all projects were executed in a controlled environment, different dependencies' versions or project configurations could still influence the results. Also, the non deterministic behavior of LLMs may cause slight variations in outputs between executions. To mitigate this, we used fixed prompts, temperature settings, and the number of executions for each LLM.

\subsubsection*{\textbf{Construct Validity}} 

A threat to construct validity arises from how migration success was defined and measured. 
Execution and coverage preservation are practical indicators of correctness but do not capture all aspects of test quality.
A migrated test may still pass while behaving differently when integrated into the full suite. To mitigate this, all migrations were executed in real project environments, using coverage as an additional quality measure to reduce false positives---cases where tests pass but fail to preserve their original intent.



\subsubsection*{\textbf{External Validity}} 

The external validity of our findings is constrained by two main factors.
First, we isolated the migration from other changes performed on the system.
Second, we analyzed a limited number of models and projects.
To mitigate these, we selected widely adopted open-source projects and validated migrations in real execution environments, ensuring that the evaluated scenarios reflect realistic development settings.

\subsubsection*{\textbf{Conclusion Validity}} 
We relied on the execution of the migrated tests to analyze the performance of LLMs.
Although all migrations were executed under controlled local environments, minor changes in dependencies or package versions could have influenced individual outcomes. 
We rebuild the projects from their verified releases and executed them multiple times to ensure stability and reproducibility of results.

\section{Related Work}

Library and framework evolution and migration are research topics largely explored by the literature in multiple ecosystems~\cite{barbosa2022, lamothe2021systematic, wang2020exploring, xavier2017historical, li2018characterising, brito2020you, hora2015developers, brito2018use, robbes2012developers, li2013does, sawant2019react, sawant2018features, nascimento2021javascript, malloy2019empirical}.
In the context of testing framework migration, Barbosa and Hora~\cite{barbosa2022} empirically explored how developers migrate Python tests from unittest to Pytest.
The authors detect that multiple popular Python projects migrated to Pytest.
In many cases, the migration was not simple, taking a long period to conclude or never concluded at all.

Recently, Large Language Models (LLMs) have been adopted in multiple software engineering tasks, including generating tests, refactoring, fixing bugs, and supporting code review~\cite{fan2023large, monteiro2023end, liang2023can, tufano2023predicting, georgsen2023beyond, hou2023large, hora2024predicting, esem2024_api_migration_llm, di2025deepmig, schafer2023empirical, alshahwan2024automated}.
Di Rocco~\emph{et al.}~proposed DeepMig, a transformer-based approach to support coupled library and code migrations in Java~\cite{di2025deepmig}.
The research presents promising results, showing that DeepMig is able to recommend both libraries and code; in several projects with a perfect match.
Almeida~\emph{et al.}~provided an initial study to explore automatic library migration using LLMs~\cite{esem2024_api_migration_llm}.
Specifically, with the support of GPT-4o, the authors migrated a client application to a newer version of SQLAlchemy, a Python Object-Relational Mapping (ORM) library.
The study presents promising results, concluding that LLMs can correctly migrate the project with only minor mistakes.
Our research contributes to the literature with a novel solution based on LLMs to support testing framework migration.

\section{Conclusion}

Our study analyzed how Large Language Models perform automated test migrations from \texttt{unittest} to \texttt{Pytest}. Across 40 isolated migrations using GPT-4o and Claude Sonnet 4, with three prompting strategies (Zero-shot, One-shot, and Chain-of-Thought) and two temperature settings (0.0 — deterministic and 1.0 — neutral), LLMs achieved an overall 48.54\% effectiveness rate.
This effectiveness was validated through real executions on 7 of the top 100 Python open-source projects on GitHub, showing that even under isolated, context-free conditions, LLM-generated migrations can maintain partial functionality in real-world software environments.

Despite this potential, several challenges remain. Both models often struggled with structural coherence, fixture adaptation, and dependency management. Claude Sonnet 4 exhibited a conservative migration style—preserving class-based architectures and legacy \texttt{unittest} references, while GPT-4o favored more transformations toward function-oriented and fixture-driven designs. Prompt strategy emerged as a key factor: One-shot and Chain-of-Thoughts improved syntactic modernization but reduced architectural fidelity. Temperature variation, however, had a negligible impact on results.


Finally, future work may explore
(i) test migrations performed by coding agents, which can perform software testing tasks autonomously~\cite{li2025rise, overmockedagentsmsr, agentminingpaper},
(ii) contribution studies~\cite{brandt2024shaken, hora2024pathspotter, danglot2019automatic} to evaluate developer acceptance by submitting LLM-migrated tests as pull requests to open-source projects,
(iii) a deeper assessment of the quality of LLM-migrated tests in comparison to manually migrated ones, and (iv) analysis on large, industry-grade repositories that rely on advanced \texttt{Pytest} features.


\section*{Acknowledgments}

This research was supported by CNPq (process 403304/2025-3), CAPES, and FAPEMIG.
This work was partially supported by INES.IA (National Institute of Science and Technology for Software Engineering Based on and for Artificial Intelligence), www.ines.org.br, CNPq grant 408817/2024-0.

\balance
\bibliographystyle{ACM-Reference-Format}
\bibliography{main}
\end{document}